\documentclass[journal=aamick,manuscript=article]{achemso}

\usepackage[usenames,dvipsnames,table]{xcolor}
\usepackage[version=3]{mhchem} %
\usepackage{miller}
\usepackage{subcaption}
\usepackage{multirow}
\usepackage{color}
\usepackage{textgreek}

\usepackage{amsmath,amssymb}
\graphicspath{{Figures/}}

\usepackage{pdfpages}

\usepackage[T1]{fontenc}
\usepackage[%
  breaklinks,             %
  bookmarks = false, %
  pdfpagemode   = UseNone,%
  pdfstartview  = FitH,     %
  pdfstartpage  = 1,      %
  colorlinks    = true,   %
]{hyperref}

\newcommand\myshade{85}
\colorlet{mylinkcolor}{YellowOrange}
\colorlet{myurlcolor}{Aquamarine}
\colorlet{mycitecolor}{violet}

\hypersetup{
  linkcolor  = mylinkcolor!\myshade!black,
  citecolor  = mycitecolor!\myshade!black,
  urlcolor   = myurlcolor!\myshade!black,
  colorlinks = true,
}



\author{Md Salman Rabbi Limon}
\affiliation{Department of Mechanical Engineering, Texas Tech University, Lubbock, Texas 79409, USA}
\author{Curtis Duffee}
\affiliation{Department of Mechanical Engineering, Texas Tech University, Lubbock, Texas 79409, USA}
\author{Zeeshan Ahmad}
\affiliation{Department of Mechanical Engineering, Texas Tech University, Lubbock, Texas 79409, USA}
\email{zeeahmad@ttu.edu}

\title{Constriction and contact impedance of ceramic solid electrolytes}


\begin{document}

\begin{abstract}

The development of solid-state batteries (SSBs) is hindered by degradation at solid-solid interfaces due to void formation and contact loss, resulting in increased impedance. Here, we systematically investigate the roles of real and unrecoverable interfacial contact areas at the electrode/\ce{Li6PS5Cl} interface in driving the impedance rise. By controlling contact geometries and applied pressures, we identify their distinct contributions to the impedance spectra and quantify their influence on the interfacial resistance and transport. Experiments reveal that interfacial resistance varies strongly with recoverable contact area and applied pressure following power law scaling, with exponents of -1 and -0.5, respectively. Moreover, distributed contacts result in lower impedance due to smaller potential gradients and a more uniform potential distribution. Continuum simulations of the contact geometries predict interfacial resistances in agreement with experiments. Our work highlights the influence of unrecoverable and recoverable contact losses on SSB impedance while quantifying the effectiveness of mitigation strategies.

\end{abstract}

Solid-state batteries (SSBs) are considered a promising next-generation energy storage technology due to their potential for improved safety, higher energy density, and the mitigation of issues such as thermal runaway.\cite{janek2016solid,zhao2020designing,wang2012thermal,feng2018thermal,finegan2024battery} However, solid-solid interfaces present unique challenges compared to solid-liquid interfaces that impede widespread adoption.\cite{janekChallengesSpeedingSolidstate2023} Interfacial phenomena such as void formation, stress-induced deformation, and contact loss lead to increased interfacial resistance and hinder ionic transport.\cite{lewisLinkingVoidInterphase2021,ahmadChemomechanicsFriendFoe2022,tianSimulationEffectContact2017a,vishnugopiAsymmetricContactLoss2023,kasemchainanCriticalStrippingCurrent2019a,sandovalStructuralElectrochemicalEvolution2023,vermaMicrostructurePressureDrivenElectrodeposition2021a} 
While external pressure can offer control over the interface, it is often limited due to practical considerations to a few MPa where considerable contact loss may be present due to unrecoverable voids~\cite{douxStackPressureConsiderations2020,zhangPressureDrivenInterfaceEvolution2020,zamanTemperaturePressureEffects2023}.
Addressing these challenges is crucial for improving the performance, longevity,  and reliability of SSBs.

Electrochemical impedance spectroscopy (EIS) offers a powerful non-destructive technique to probe interfacial contact in SSBs.\cite{zhang2024recent}
By analyzing frequency-dependent responses, EIS enables the separation of different impedance contributions, such as bulk transport, grain boundary effects, and interfacial processes. This capability makes it especially useful for monitoring changes in contact at the solid-solid interfaces during operation. The differences in EIS at high and low frequencies due to non-ideal contacts were elucidated by Fleig and Maier\cite{fleigFiniteElementCalculations1996,fleigRoughElectrodesSolid1997} who found that constriction effects are significant at low frequencies but disappear at high frequencies. Signatures of interfacial roughness have also been found in the EIS~\cite{delevieImpedanceElectrodesRough1989,kantSituElectrochemicalImpedance2021,delevieInfluenceSurfaceRoughness1965}. 
Recently, Eckhardt et al.\cite{eckhardtInterplayDynamicConstriction2022,eckhardt3DImpedanceModeling2022b} showed that conventional 1D equivalent circuit models are inadequate for modeling the dynamic constriction caused by contact in SSBs. Hence, interpreting experimental EIS data is challenging due to the interplay of geometric and electrochemical effects at the interface. Constriction or contact resistance has been difficult to isolate from signatures of other phenomena in EIS data since interfacial contact cannot be directly controlled in experiments~\cite{sandovalStructuralElectrochemicalEvolution2023}.

In this work, we combine EIS experiments and continuum simulations to investigate the effect of contact morphology at the electrode/\ce{Li6PS5Cl} solid electrolyte (SE) interface on the impedance by systematically controlling the recoverable and unrecoverable contact loss in experiments. We precisely control the unrecoverable contact loss by blocking transport in certain regions of the interface. We find distinct signatures of interfacial contact state in EIS quantified through the interfacial resistance. 
We map out the dependence of the interfacial resistance on both the unrecoverable and real contact area controlled by pressure. 
Through measurements over a wide pressure range, we find that the interfacial resistance generally scales as $P^{-0.5}$ in agreement with electrical contact resistance theories.\cite{slade2017electrical}. Further, the interfacial resistance rises rapidly as the unrecoverable contact area increases following a power law scaling of -1. The interfacial resistance can be approximated by the expression for constriction resistance~\cite{newmanResistanceFlowCurrent1966,holmElectricContacts1967,greenwoodConstrictionResistanceReal1966} for small contact areas. For the same interfacial contact area, we show that a contact geometry that is more evenly distributed results in lower impedance compared to a concentrated one  due to a more uniform potential distribution.

To complement the experiments, we implemented a finite-element model that incorporates electromagnetic contact conditions and contact loss at the electrode/SE interface. We simulate the experimental geometry using this model which demonstrates strong agreement with the measured EIS and interfacial resistance. Our simulations provide insights into the spatial distribution of potential gradients and their dependence on contact geometry.

\begin{figure}[htbp]
    \centering
    \includegraphics[width=0.6\textwidth]{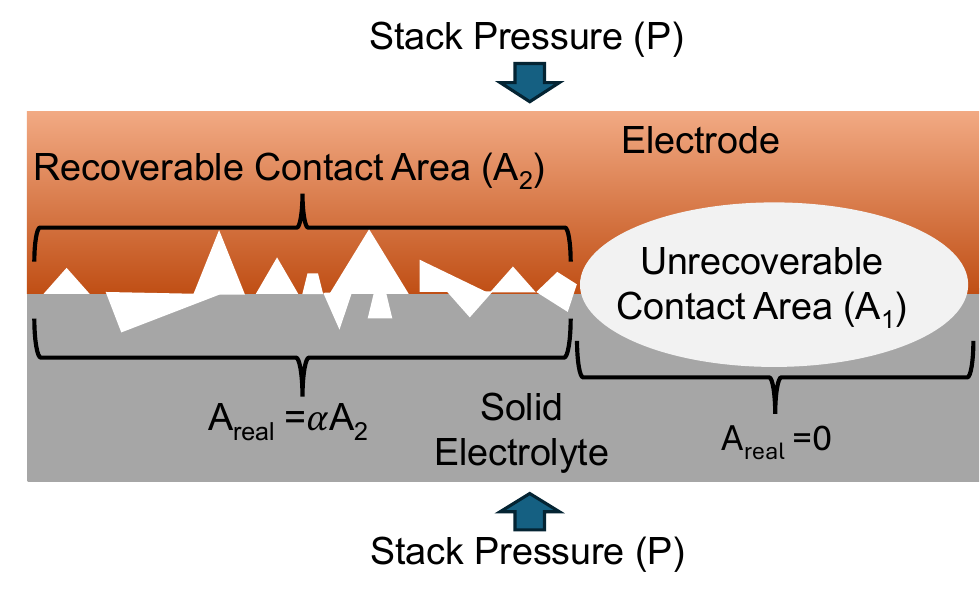}
    \caption{Schematic of the electrode-SE interface under stack pressure (P). The interface is divided into recoverable contact area (\(A_2\)) and unrecoverable contact area (\(A_1\)). The recoverable area contributes to the real contact area (\(A_{\text{real}} = \alpha A_2\)), which increases with applied stack pressure, while the unrecoverable area does not contribute to contact ($A_{\text{real}} = 0$) and is independent of pressure.
}
    \label{fig:contact}
\end{figure}

We consider the schematic electrode-SE interface in \autoref{fig:contact} and divide it into two regions: recoverable and unrecoverable contact areas. The recoverable region refers to areas where contact can be reestablished or adjusted by applying pressure, allowing for modulation of the contact morphology. The unrecoverable region, on the other hand, represents areas where no contact occurs, and the morphology remains unaffected by pressure. This distinction is critical for understanding and optimizing interfacial contact in SSBs as unrecoverable contact areas involve large interfacial voids which may require pressures much higher than practically feasible values to establish contact\cite{zhangPressureDrivenInterfaceEvolution2020,zamanTemperaturePressureEffects2023}. Measures other than external pressure may be required to mitigate unrecoverable contact losses. We define $\tilde{\gamma}$ as the fraction of recoverable (nominal) contact area and $\alpha$ as the fraction of real contact area in recoverable regions. $\alpha$ depends on pressure and surface roughness. $\gamma$ represents the overall fraction of the real contact area with respect to the total nominal area. In the example of \autoref{fig:contact}, $\tilde{\gamma}=A_2/(A_1+A_2)$, and $\gamma=\alpha A_2 /(A_1+A_2)$.

\noindent \textit{Experimental}. \ce{Li6PS5Cl} powder was pressed into pellets using a 12 mm diameter die and a pressure of 375 MPa for 2 minutes.\cite{sandovalStructuralElectrochemicalEvolution2023}
This high-pressure fabrication step was employed to reduce porosity and enhance ionic conductivity.\cite{douxPressureEffectsSulfide2020,sakuda2013sulfide,hayashi2016development}
A thin paper (thickness: 55 $\mu$m) was used during the experiments to control the unrecoverable contact area between the pellet and the steel electrode by acting as an insulating material to block transport at the interface. Experiments were conducted to analyze the effect of varying recoverable contact area, real contact area (through pressure), and geometrical distribution of contact loss for the same total contact area. After placing the paper on the pellet, a pressure of 180 MPa was applied to establish contact. Following the EIS tests for a specific contact area, the paper was removed, and the bare pellet was re-pressed at the fabrication pressure to ensure consistent pellet conditions for subsequent tests. Two different pellets were used for experiments involving variation of recoverable contact area and geometrical distribution of contact.

\noindent \textit{Simulations}. Finite-element simulations were conducted to solve for the complex potential, $\phi=\phi_r+ i \phi_i$ using the equation $\nabla^2 \phi = 0$ at steady state following \citet{fleigFiniteElementCalculations1996}. We used the Dirichlet boundary condition for potential at the metal/SE interface and the Neumann boundary condition at the free sides of the SE. Across the metal/insulator interface, the normal component of the current density $j$ was set to be equal. 
Simulations were performed at different frequencies using the expression for complex conductivity $\kappa = \sigma + \mathrm{i} \omega \epsilon$ where $\sigma$ is the conductivity of the SE, $\epsilon$ is the permittivity, and $\omega$ is the angular frequency. To obtain the EIS, the complex impedance was calculated using $Z(\omega)  = \Delta \phi /\int j dS$ where $\Delta \phi$ is the potential difference between the electrodes and $j= -\kappa \nabla \phi $ is the current density perpendicular to the area $dS$. 
The model assumes perfect contact in recoverable regions without microscopic roughness, hence, experimental results at the highest pressure were only compared with the simulations.
The equations were solved using the open-source Multiphysics Object-Oriented Simulation Environment (MOOSE) framework~\cite{Gaston2009moose}. Code for reproducing the simulations is available on GitHub~\cite{arg2024contactechem}.

\begin{figure}[htbp]
    \centering
    \includegraphics[width=\textwidth]{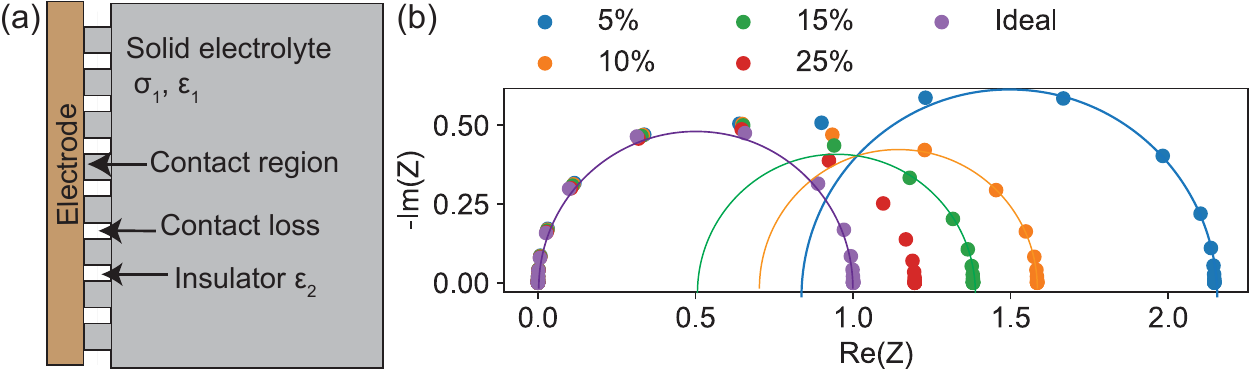}
    \caption{2D model for electrode/SE interface in the presence of contact loss used to calculate EIS. (a) Schematic of simplified contact geometry characterized by a rough grooved surface at the electrode/SE interface. (b) Comparison of EIS of the system with different recoverable contact area fractions $\tilde{\gamma}$ showing the impact on the low frequency semicircle.}
    \label{fig:toy}
\end{figure}

We first demonstrate the effect of contact loss on EIS through a simple 2D geometry of electrode/SE interface shown in \autoref{fig:toy}a. The interface has gaps whose widths were varied to control $\tilde{\gamma}$. 
\autoref{fig:toy}b shows the EIS of the geometry considered in (a) as a function of the $\tilde{\gamma}$. A striking difference emerges between the EIS for ideal contact ($\tilde{\gamma}=100\%$) and non-ideal contact ($\tilde{\gamma} < 100\%$). Ideal contact yields a single semicircle in the EIS, whereas non-ideal contact introduces an additional semicircle, whose size increases with decreasing $\tilde{\gamma}$. Notably, the high-frequency semicircle remains invariant with $\tilde{\gamma}$, indicating its dependence on the bulk properties of the SE. In contrast, the low-frequency semicircle exhibits a strong sensitivity to interfacial contact, hence we attribute it to interfacial impedance~\cite{eckhardtInterplayDynamicConstriction2022,fleigFiniteElementCalculations1996}.

\begin{figure}[htbp]
    \centering
    \includegraphics[width=\textwidth]{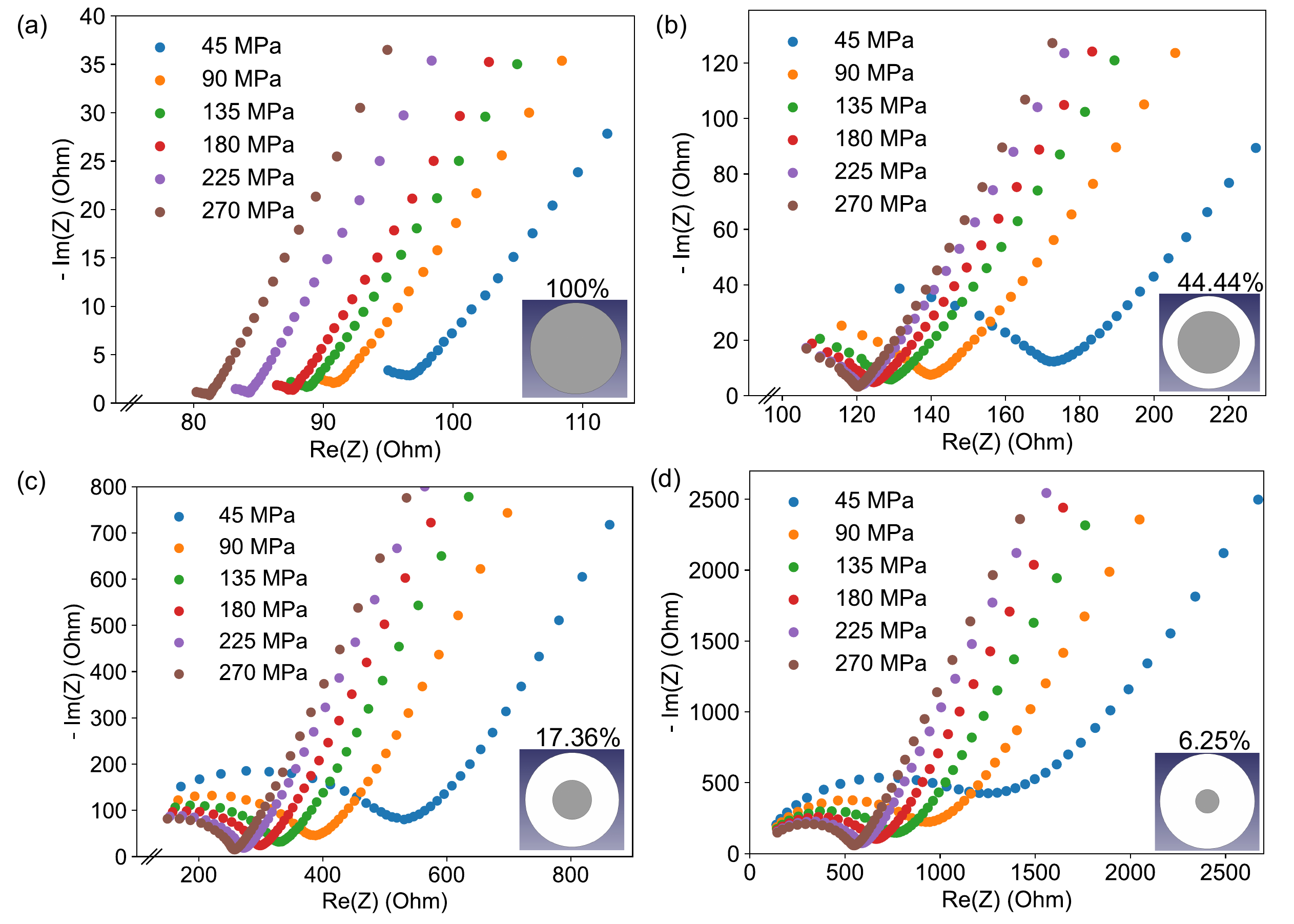}
    \caption{EIS of a \ce{Li6PS5Cl} pellet at different applied pressures (45–270 MPa) and recoverable contact areas ($\tilde{\gamma}$). The measurements were taken at four different values of $\tilde{\gamma}$: (a) 100\% (full contact with a 12 mm diameter pellet), (b) 44.44\% (one side with an 8 mm diameter contact), (c) 17.36\% (one side with a 5 mm diameter contact), and (d) 6.25\% (one side with a 3 mm diameter contact). The gray-shaded regions of the insets illustrate the contact areas for each case. 
    Both higher pressures and larger contact areas reduce the impedance. }
    \label{fig:nyq-exp}
\end{figure}

We use the insights from this 2D model to deconvolute the contributions from the bulk and interfacial contact to the experimentally measured impedance of \ce{Li6PS5Cl}-steel electrode interfaces under different applied pressures and $\tilde{\gamma}$. \autoref{fig:nyq-exp} illustrates the impact of applied pressure on EIS of \ce{Li6PS5Cl} SE for four different values of $\tilde{\gamma}$: (a) $100\%$, (b) $44.44\%$, (c) $17.36\%$, and (d) $6.25\%$. 
The configurations with different $\tilde{\gamma}$ were generated by blocking transport with paper between the steel electrodes and the pellet in the white regions shown in the insets of \autoref{fig:nyq-exp} so that the gray shaded area corresponds to contact region with diameters of 12 mm, 8 mm, 5 mm, and 3 mm. For each $\tilde{\gamma}$, the applied pressure controls the real area of contact fraction $\gamma$, i.e. higher pressure increases $\gamma$.
As expected, higher pressures and $\tilde{\gamma}$ reduce the impedance.  
The EIS data were fitted using an equivalent circuit consisting of a resistor (R1) in series with a parallel combination of a resistor (R2) and a constant phase element (CPE), followed by another CPE in series (Fig. S3). From the fitting, it was observed that R1 is generally not sensitive to changes in $\tilde{\gamma}$ and pressure (Fig. S4) while R2 varies significantly. Hence, R1 can be attributed to the bulk resistance (\(R_{\text{bulk}}\)) of the material, while R2 can be assigned to the interfacial resistance (\(R_{\text{int}}\)). From  the EIS data at $\tilde{\gamma}=1$, we obtain the bulk conductivity, $\sigma_b$ of \ce{Li6PS5Cl} as 2.947 mS/cm.

\begin{figure}[htbp]
    \centering
    \includegraphics[width=\textwidth]{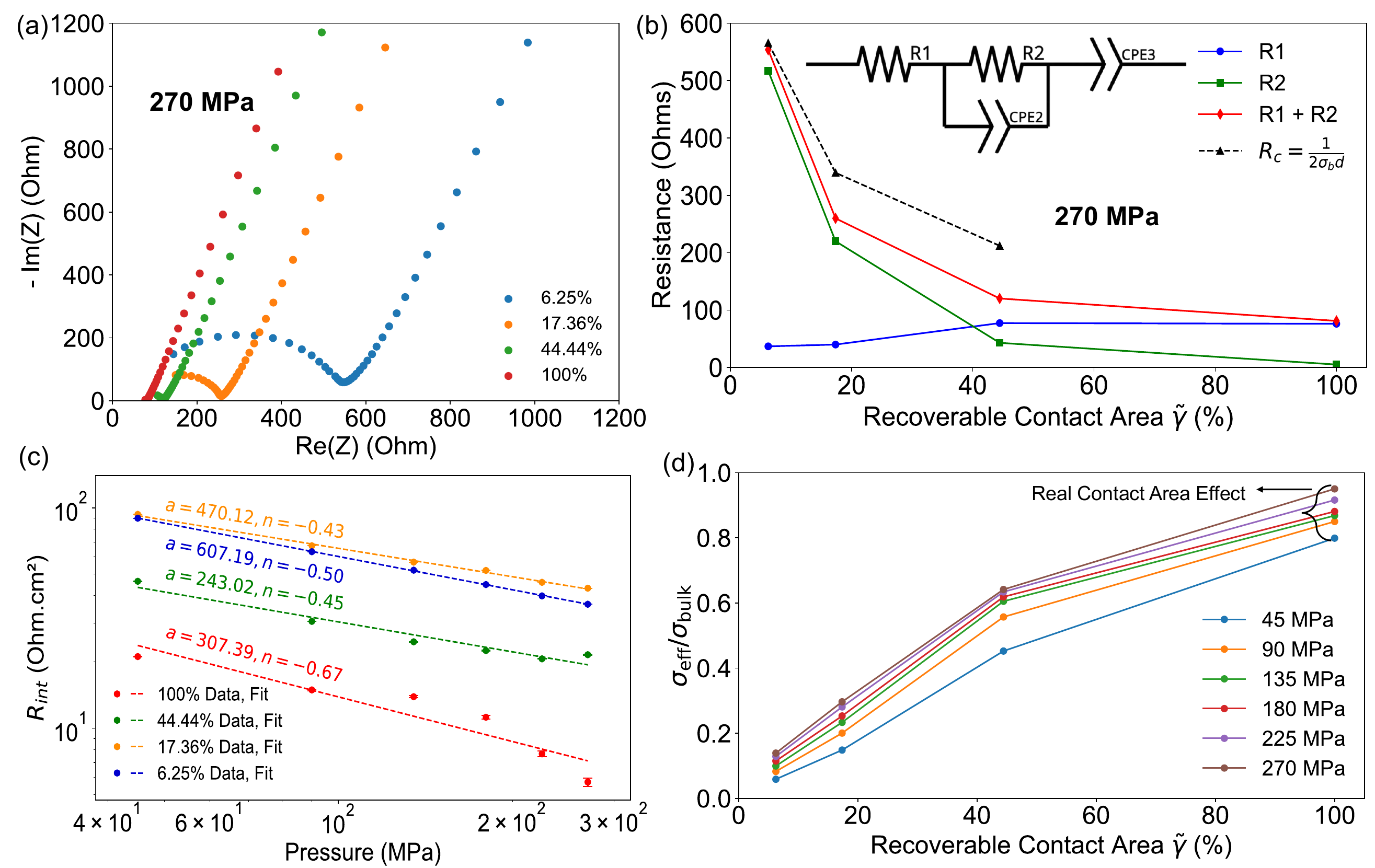}
    \caption{Analysis of interfacial resistance and effective conductivity of the pellet with contact loss and pressure.
    (a) EIS plots with different $\tilde{\gamma}$ at 270 MPa, the highest pressure considered, assumed to represent ideal contact conditions. (b) Fitted resistance values as a function of contact area at 270 MPa, using the equivalent circuit model shown in the inset. The bulk resistance (R1) remains constant but the interfacial resistance (R2) increases with a decrease in $\tilde{\gamma}$. The black dotted line plots the constriction resistance for $\tilde{\gamma}<1$, $R_c = 1/(2d\sigma_b)$ derived based on the bulk conductivity ($\sigma_b$) and diameter of contact $d$.
    (c) Interfacial resistance ($R_{\text{int}}$) as a function of applied pressure on a log-log scale for different $\tilde{\gamma}$. The data points were fitted to the power-law equation, $R_{\text{int}} = aP^n$. 
    (d) Variation of the normalized effective conductivity ($\sigma_{\mathrm{eff}} / \sigma_{\mathrm{bulk}}$) with $\tilde{\gamma}$ showing an increase with  both pressure and $\tilde{\gamma}$.
    At 100\% contact and 270 MPa pressure, the normalized conductivity approaches its maximum value. The curly bracket at $\tilde{\gamma}=1$ indicates the potential for increasing effective conductivity through pressure which determines the real area of contact.}
    \label{fig:eff_con}
\end{figure}

In \autoref{fig:eff_con}, we analyze the fitted equivalent circuit parameters and effective conductivity of \ce{Li6PS5Cl} under different pressures and contact conditions.  Panel (a) presents the EIS plots for different $\tilde{\gamma}$ at 270 MPa. The intercept on the real axis, indicative of total resistance, decreases as the contact area increases. Panel (b) shows the fitted bulk, interfacial, and total resistances as a function of $\tilde{\gamma}$ at 270 MPa. At high $\tilde{\gamma}$, lack of sufficient high frequency data points led to higher errors in the fitting parameters.
The interfacial resistance rises rapidly with a decrease in $\tilde{\gamma}$ and
follows the power law scaling $R_{\text{int}} = a \tilde{\gamma}^n$ with an exponent $n$ equal to  -1 (Fig. 
 S6). We also plot the constriction or spreading resistance $R_c=1/(2d\sigma_{b})$ where $d$ is the diameter of circular contact area and $\sigma_{b}$ is the bulk conductivity. This expression corresponds to a power law scaling with exponent -0.5 and is commonly used in literature to incorporate constriction effects~\cite{mcconohyMechanicalRegulationLithium2023a,singhNonLinearKineticsLithium,krauskopfFastChargeTransfer2020}. We find that this expression generally overestimates the interfacial resistance but becomes more accurate as the diameter of the contact decreases (e.g., for microelectrodes). However, it should be noted that the contact area estimated from constriction resistance expression does not equal the real area of contact and  gives only an upper estimate~\cite{greenwoodConstrictionResistanceReal1966}.

A fundamental question in the theory of electrical contacts is how the real area of contact and resulting interfacial resistance scale with pressure~\cite{holmElectricContacts1967,slade2017electrical,llewellyn-jonesPhysicsElectricalContacts1957}. As the pressure is increased, the real area of contact increases, thus reducing the interfacial resistance. 
The $R_{\text{int}}$ vs. $P$ scaling provides insights on the nature of deformation at the the interface, current constriction, asperity interactions, and  real contact area~\cite{greenwoodConstrictionResistanceReal1966,krauskopfFundamentalUnderstandingLithium2019,zhangPressureDrivenInterfaceEvolution2020}. For plastic deformation, the constriction resistance scales as $P^{-1/2}$~\cite{llewellyn-jonesPhysicsElectricalContacts1957}.
At the microscopic level, Hertzian contact theory predicts $A_{\text{real}} \propto P^{2/3} $, hence $R_{\text{int}}\propto P^{-2/3}$ if the contact at each asperity can be considered in parallel ($R^{-1}=\Sigma R_i^{-1}$) and asperity interactions are ignored~\cite{greenwoodConstrictionResistanceReal1966,yeoEffectAsperityInteractions2010}. 

We plot log $R_{\text{int}}$ vs. log P  in \autoref{fig:eff_con}c for different $\tilde{\gamma}$ and fit the data to a power-law equation, \( R_{\text{int}} = aP^n \).
We find that the exponent $n$ is nearly -0.5 for all $\tilde{\gamma}$ except for $\tilde{\gamma}=1$ where it is -0.67. The value of -0.5 is consistent with other experiments and theory of constriction resistance in electrical contacts, indicating that plastic deformation may be occuring at the interface due to high stresses encountered at the asperities\cite{slade2017electrical}.
Our results for the steel-SE interface agree with reported value for the Li-SE interface by \citet{krauskopfFundamentalUnderstandingLithium2019} but not -1 obtained by \citet{zhangPressureDrivenInterfaceEvolution2020}.  
The different behavior of $R_{\text{int}}$ at $\tilde{\gamma}=1$ warrants further investigation into the role of unrecoverable contact area.

A useful metric for SEs for adoption in practical SSBs is the effective conductivity that incorporates the effects of various factors such as contact loss and tortuosity and represents the real rate capability of the battery. An effective conductivity of several mS/cm is required for SSBs to compete with liquid-electrolyte based batteries~\cite{janekChallengesSpeedingSolidstate2023}. The most common cause of contact loss in SSBs is the formation of voids at the metal anode/SE interface during stripping~\cite{kasemchainanCriticalStrippingCurrent2019a,ahmadChemomechanicsFriendFoe2022}.  Further, during battery cycling, mechanical and electrochemical stresses can generate cracks that cause contact loss, leading to an increase in interfacial resistance and decrease in effective conductivity. We calculate this effective SE conductivity using $\sigma_{\text{eff}} = l/(R_{\text{tot}}A)$ where $l$ and $A$ are the pellet thickness and total area. 
\autoref{fig:eff_con}d plots the normalized $\sigma_{\text{eff}}$ at different applied pressures and $\tilde{\gamma}$.  We find a much higher sensitivity of $\sigma_{\text{eff}}$ to $\tilde{\gamma}$ compared to pressure. When the contact area falls below 50\%, there is a steep decrease in $\sigma_{\text{eff}}$, which will drastically impair the SSB rate capability. Hence, it is important to avoid macroscopic contact losses caused by large unrecoverable voids at the electrode-SE interface.  These findings highlight the importance of sustaining adequate contact area during repeated cycling to optimize the electrode/SE interface and maintain efficient transport in SSBs.

\begin{figure}[htbp]
    \centering
    \includegraphics[width=\textwidth]{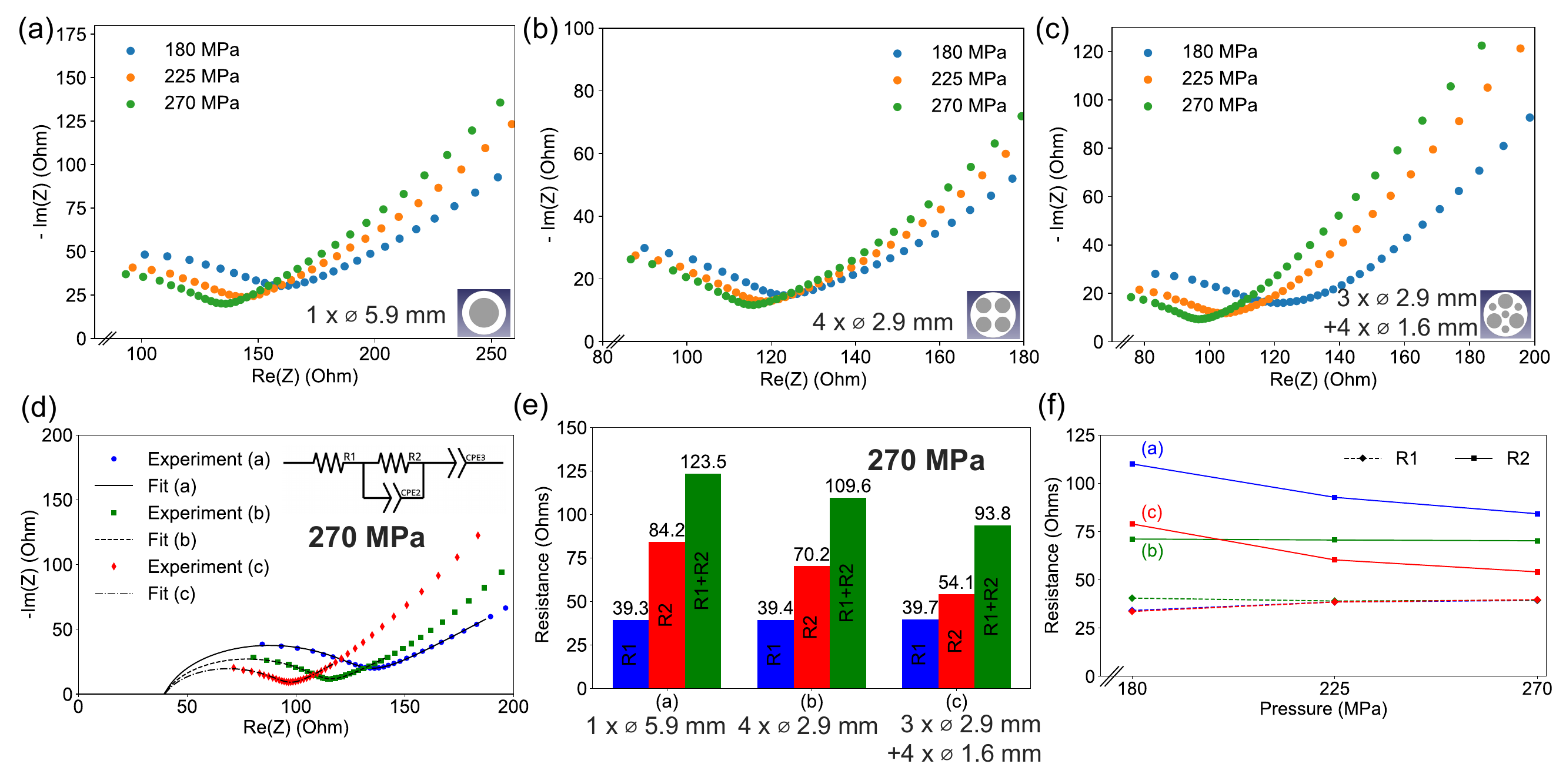}
    \caption{EIS analysis of interfacial resistance as a function of contact geometry and pressure. Nyquist plots for EIS data measured at  180 MPa, 225 MPa, and 270 MPa with three different contact geometries: (a) a single circular contact with a 5.9 mm diameter, (b) four contacts with a 2.95 mm diameter each, and (c) three contacts with a 2.90 mm diameter combined with four contacts with a 1.55 mm diameter as shown in the insets. All three configurations have the same total contact area. (d) Comparison of experimental and fitted data at 270 MPa for the three cases using the equivalent circuit shown in the inset, illustrating that the bulk resistance is constant across configurations. (e) Variation of bulk resistance (R1) and interfacial resistance (R2) at 270 MPa with contact geometries, showing that with more distributed contacts, the interfacial resistance (R2) decreases significantly. (f) Variation of bulk (R1) and interfacial resistance (R2) as a function of applied pressure for the three contact geometries.}
    \label{fig:same-A}
\end{figure}

The above experiments quantified contact loss effects while keeping similar contact geometries: a single circular area at the center of the pellet. Next, we study the relationship between the location and distribution of contact loss and the interfacial impedance while keeping $\tilde{\gamma}$ constant.  This study is motivated by the work of \citet{greenwoodConstrictionResistanceReal1966} who obtained the resistance for different locations of microcontacts. 
\autoref{fig:same-A} compares the impedance due to contact loss with the same value of $\tilde{\gamma}=$ 24.2\%  for three different contact configurations: (a) a single circular contact with a 5.9 mm diameter ($1\times \phi 5.9$ mm), (b) four circular contacts with a 2.95 mm diameter each ($4\times \phi 2.95$ mm), and (c) a combination of three 2.9 mm diameter contacts and four 1.55 mm diameter contacts ($3\times \phi 2.9 + 4 \times \phi 1.55$ mm) as shown in the insets at pressures of 180 MPa, 225 MPa, and 270 MPa. 
Panel (d) shows the equivalent circuit fit while (e) compares the values of resistance extracted from the fits for 270 MPa pressure. 
Panel (f) plots the bulk and interfacial resistances for the three configurations against pressure. Across all geometries, increasing pressure reduces the interfacial resistance while the bulk resistance remains nearly constant. The interfacial resistance is sensitive to the distribution of contact and non-contact regions. 
The geometry with the most evenly distributed contact, configuration (c), exhibits the lowest interfacial resistance, 36\% lower than (a). We believe that the lower interfacial resistance is due to smaller potential gradients and more uniform current density when the contact loss is more evenly distributed.

\begin{figure}[htbp]
    \centering
    \includegraphics[width=\textwidth]{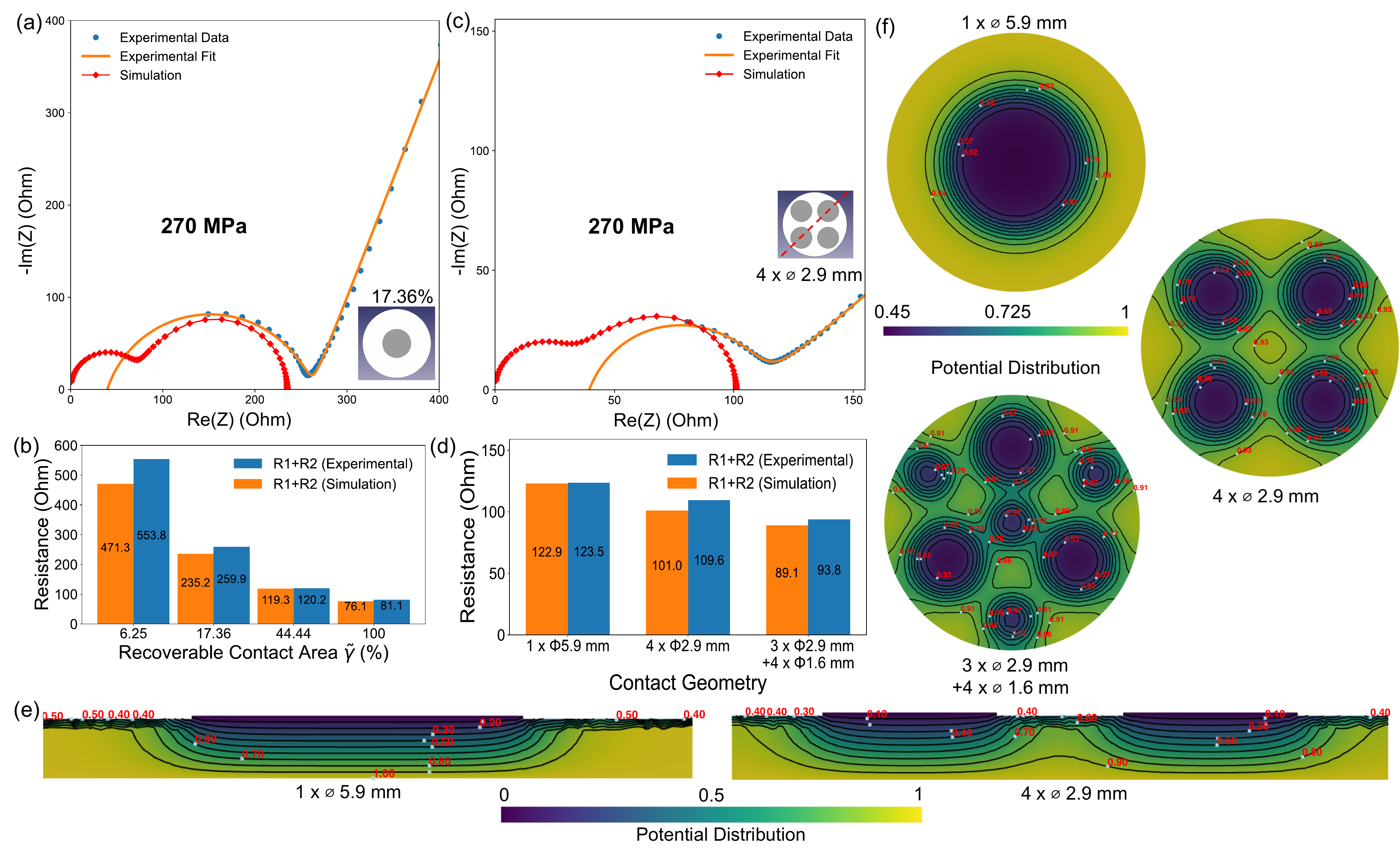}
    \caption{Comparison of experimental and simulation results for different contact areas and geometry-dependent interfacial resistance. (a) Representative EIS plot comparing experimental and simulated EIS for $\tilde{\gamma} = 17.36\%$. (b) Comparison of total resistance (R1 + R2) between experiment and simulation for different $\tilde{\gamma}$, demonstrating good agreement across a wide range. (c) Representative EIS plot comparing experimental and simulated EIS for the same total contact area but different geometries ($4\times \phi2.95$ mm configuration).
    (d) Comparison of total resistance (R1+R2) between experiment and simulation for different geometries with the same total contact area, confirming the model's ability to capture the effect of geometry on interfacial resistance. For both experimental and simulation results, the total resistance decreases as the contact becomes more distributed.
    (e) Potential distribution contours (side view cross-section) for two different contact geometries with the same $\tilde{\gamma}$: $1\times \phi5.9$ mm and $4\times \phi2.95$ mm. The cross sections are taken through the centers of the circular contacts, as shown in the inset of (c). 
    (f) Potential distribution contours (top view cross-section) for the three geometries with the same $\tilde{\gamma}$ at a distance of 0.52 mm from the top surface with $\phi=0$ V. %
    All potential distributions are plotted at a frequency of 726.124 kHz. 
    The potential distribution (e-f) reveals localized variations in the contact regions, highlighting the influence of contact geometry. The single contact configuration shows a more localized potential drop, while configurations with multiple smaller contacts exhibit lower potential gradients, leading to lower resistance.
    }
    \label{fig:exp-sim}
\end{figure}

To gain insights into the effects of constriction and contact loss for the geometries considered in the experiments, we apply the developed model to simulate the experimental contact geometries and predict the current and potential distribution for EIS. 
\autoref{fig:exp-sim}a and c compare the representative EIS plots obtained from the simulation with the experimental data for the contact geometries shown in the insets. The corresponding equivalent circuit parameters are compared in panels (b) and (d). It is worth noting that the simulation is  able to replicate the experimental EIS solely by incorporating the contact geometry, indicating that contact loss is the predominant factor influencing the impedance.
The slight underestimation of the resistance may be due to factors not considered in the simulations such as applied pressure, roughness/asperities, surface layers and impurities, etc. The total resistance decreases with $\tilde{\gamma}$ according to both simulations and experiments. Further, the resistances for the three configurations with the same value of $\tilde{\gamma}$ are plotted in panel (d), showing that the configuration with distributed contact loss exhibits the lowest resistance.

\autoref{fig:exp-sim}e plots the color map and contours for the absolute value of the complex potential for the contact configurations considered for experiments with the same recoverable area $\tilde{\gamma}=24.2\%$ using both  side and top cross-sectional views at a frequency of 726.12 kHz.
The side view cross sectional plane for $1\times \phi5.9$ mm configuration passes through the center of the single circular contact with contact loss regions at the ends while for $4\times \phi 2.95$ mm configuration, the plane passes diagonally through the centers of two circular contacts and the center of the pellet as shown by the red line in panel (c)'s inset. 
Although constriction effects on potential contours are expected to diminish at high frequencies~\cite{fleigFiniteElementCalculations1996}, the equipotential lines indicate that they they persist even at this high frequency. 
We observe a high density of potential contours in the non-contact regions, indicating a large potential drop. However, the drop is higher for the $1\times \phi5.9$ mm configuration compared to the $4\times \phi 2.95$ mm one where the potential contours merge, leading to a more uniform potential distribution.
The uniformity of the potential distribution can be visualized through the top views of cross-sectional area for the three configurations taken at the same distance $=0.52$ mm from the top electrode shown in \autoref{fig:exp-sim}f. For $1 \times \phi5.9$ mm configuration, the low potential region is very localized, leading to uneven distribution and higher resistance due to current focusing. In contrast, the configurations with multiple contact regions exhibit increasingly uniform potential distributions, minimizing localized potential gradients, allowing current to flow more uniformly and efficiently across the interface, resulting in lower resistance. Simulations also reveal that contact loss near the center of the pellet is more detrimental compared to the outer region (Fig. S9).

To summarize, our work demonstrates that the real and unrecoverable contact losses as well as the geometry of contacts critically affect the interfacial resistance and effective ionic conductivity of SEs.
Increasing the recoverable contact area, applied pressure, and distributing the contact area evenly were found to minimize the interfacial resistance. Our experiments reveal power law scaling for variation of interfacial resistance with pressure and recoverable contact with exponents -0.5 and -1 respectively. Our simulations accurately capture the interfacial resistance due to variations in the recoverable contact area and its distribution obtained from experiments. Our findings, which quantify the variation in interfacial resistance, offer fundamental insights into the mechanisms behind the impedance rise in SSBs due to interfacial contact loss.

\begin{acknowledgement}
We acknowledge the Texas Tech University Mechanical Engineering Department startup grant for support of this research. C. D. acknowledges support from the Honors College at Texas Tech University. We acknowledge the High-Performance Computing Center (HPCC) at Texas Tech University and the Lonestar6 research allocation (DMR23017) at the Texas Advanced Computing Center (TACC) for providing computational resources that have contributed to the research results reported within this paper.

\end{acknowledgement}
\begin{suppinfo}

Details of experimental procedures and finite-element simulations, additional plots and contours from experiments and simulations. 

\end{suppinfo}

\bibliography{zotero_refs,refs}


\section*{Supplementary material}

# Supporting Information: Constriction and contact impedance of ceramic solid electrolytes

Md Salman Rabbi Limon, Curtis Duffee, and Zeeshan Ahmad[*]

Department of Mechanical Engineering, Texas Tech University, Lubbock, Texas 79409, USA

E-mail: zeeahmad@ttu.edu

# Experimental details

$Li_6PS_5Cl$ powder for preparing the pellets was purchased from NEI Corporation. The pellets were annealed at 550°C for 4 hours, a temperature known to optimize ionic conductivity and reduce activation energy.[1,2] To ensure optimal electrode-electrolyte contact, stainless steel (SS) electrodes were polished to a mirror-like finish using 5000-grit sandpaper. Due to the highly hygroscopic nature of $Li_6PS_5Cl$, all pellet preparation and experiments were conducted in an argon-filled glovebox ($H_2O$ and $O_2$ levels $\leq$ 0.1 ppm) to minimize air and moisture exposure. To maintain airtight conditions, O-rings were placed on both sides of the SS electrodes. To prevent performance degradation caused by aging effects, all experiments were carried out within 3 hours after sintering, as studies indicate that impedance growth can begin as early as the next day, even under argon-filled conditions.[3]

**Pellet Specifications:** Two types of pellets were prepared for experiments and simulations:

1. **Pellet for Different Contact Area Experiments:**

Thickness: 2.57 mm

Diameter: 12 mm

Conductivity: 0.280 S/m (Measured by EIS at 270 MPa for the bare pellet.)

2. **Pellet for Same Area with Varying Geometric Distributions:**

Thickness: 1.12 mm

Diameter: 12 mm

Conductivity: 0.248 S/m (Measured by EIS at 270 MPa for the bare pellet.)

Relative density of the pellets = 92.51% [Based on experimental lattice parameters of cubic $Li_6PS_5Cl$, Space group: F-43m, a=9.8290 Å][4]

For both cases, the simulation parameters, including dimensions and conductivity, were aligned with the experimental conditions. For the simulations, a relative permittivity of 1.4[5] was used for the paper and 5.0[6] for $Li_6PS_5Cl$.

**Details of heat treatment of $Li_6PS_5Cl$ pellets:** The $Li_6PS_5Cl$ pellets underwent

a comprehensive heat treatment process comprising drying and annealing phases to ensure optimal material properties. Initially, the material was dried to remove volatile content by gradually increasing the temperature from room temperature to 120°C over 60 minutes, holding at 120°C for 60 minutes, and cooling back to room temperature over another 60 minutes. Following this, the material was annealed through a controlled baking process. The temperature was ramped from 0°C to 120°C in 60 minutes, from 120°C to 300°C in 120 minutes, and from 300°C to 550°C in 125 minutes. It was held at 550°C for 240 minutes to enhance crystallinity and relieve internal stresses, followed by a controlled cooling process: from 550°C to 300°C over 125 minutes and from 300°C to room temperature over 150 minutes.

**EIS Setup**

All the potentiostatic EIS measurements were performed on a BioLogic SP-200 potentiostat at room temperature with a frequency range of 7 MHz to 1 Hz. The amplitude of the applied voltage was 10 mV. The pressure-dependent data for various contact areas were recorded under pressures ranging from 45 MPa to 270 MPa, controlled precisely using a hydraulic press. For the bare pellet configuration, the pellet maintained full macroscopic contact with the stainless steel (SS) electrodes. In all other configurations, an insulating layer of minimal thickness (55 $\mu$m) was applied to cover the annular region, as illustrated in the insets of Figure 3 (a-d) of the main paper. This setup allowed the central circular area to remain in direct contact with the electrodes, allowing for controlled variation in the contact area. To examine the impact of contact geometry on interfacial resistance while maintaining a constant total contact area($\sim$ 24.2% of the pellet's one circular surface ) three configurations were tested: a single circular contact (5.9 mm diameter), four smaller contacts (2.95 mm diameter each), and a mixed geometry of three contacts (2.90 mm diameter) combined with four smaller contacts (1.55 mm diameter). For the fitting of the EIS data we used impedance.py[7] package.

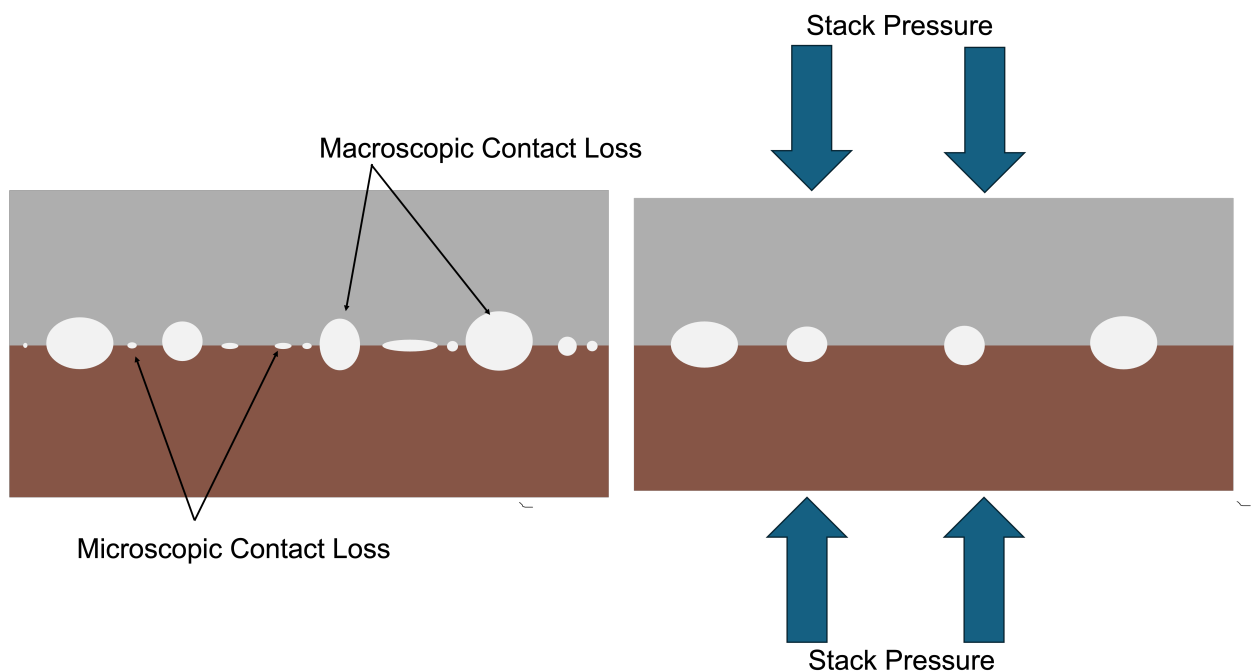

Figure S1: Effect of stack pressure on interfacial voids. Under applied stack pressure, microscopic voids at the interface are eliminated, improving contact at the microscopic level. However, larger macroscopic voids persist despite the application of pressure.

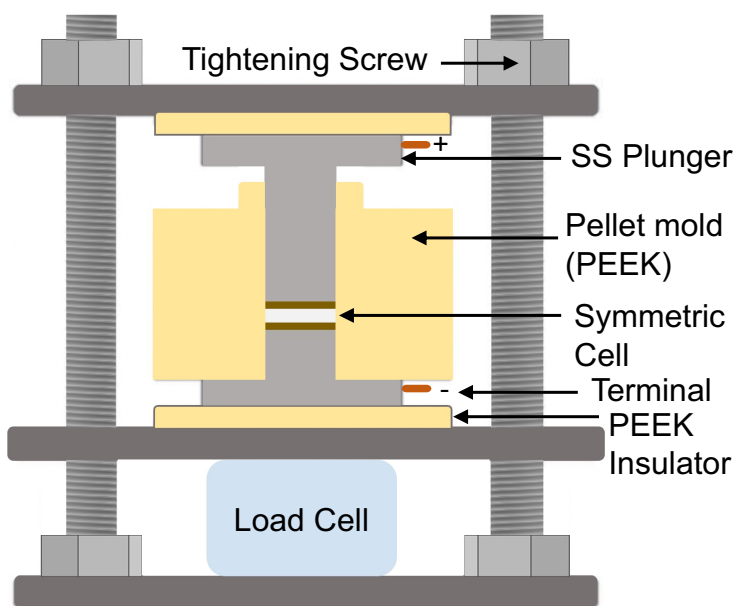

Figure S2: Schematic of pressure-controlled coin cell to measure EIS.

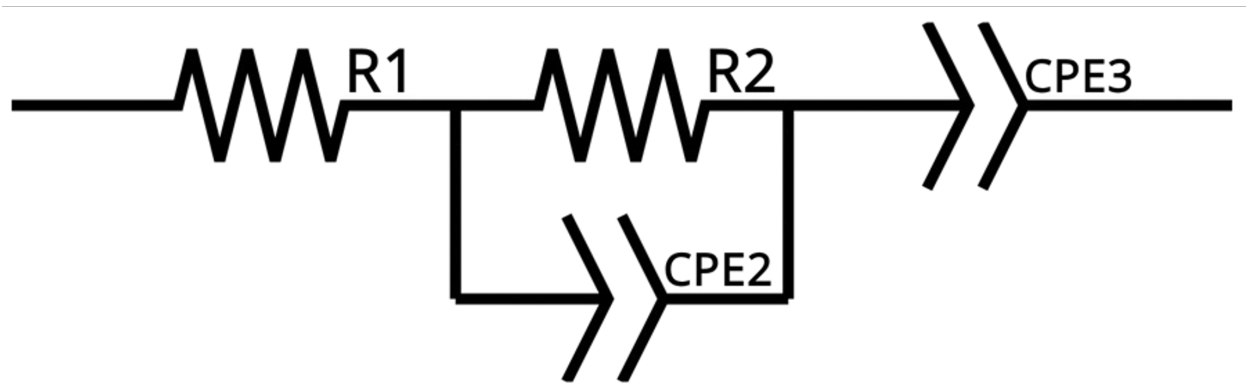

Figure S3: Equivalent circuit used for fitting the EIS data.

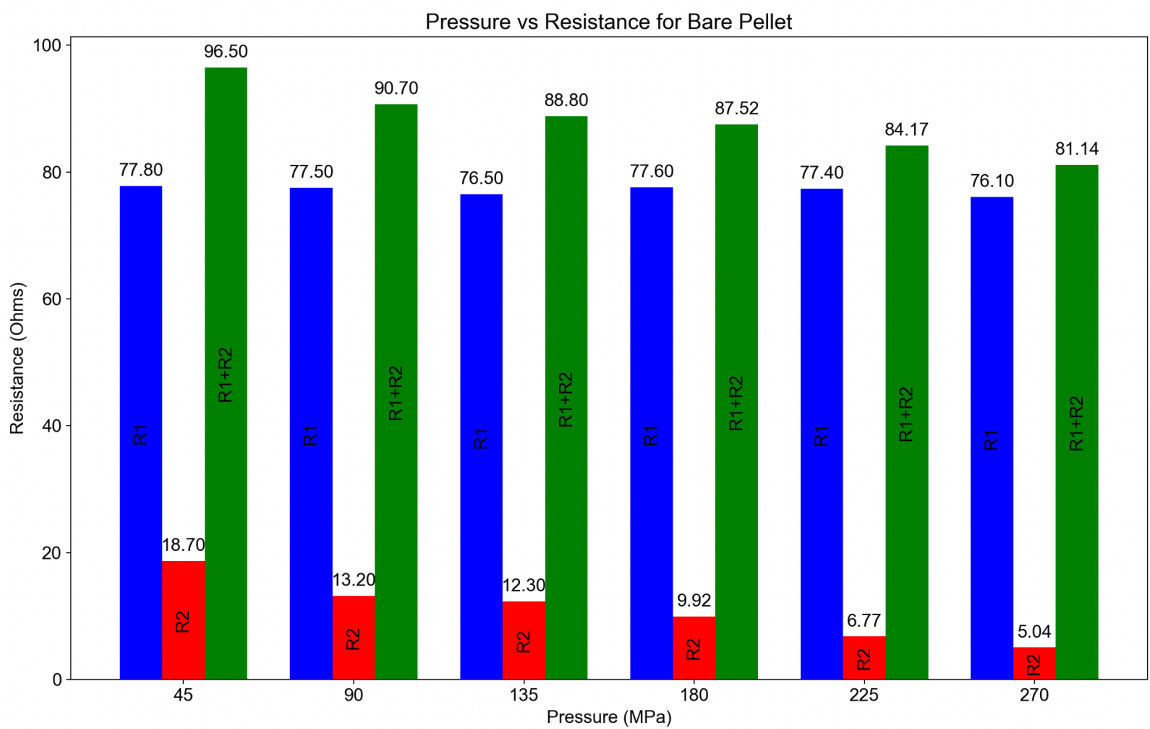

Figure S4: Resistance components (R1 and R2) of a bare pellet as a function of applied pressure (45–270 MPa). The total resistance (R1+R2) decreases with increasing pressure. Bulk resistance (R1) remains constant with pressure, while interface resistance (R2) decreases significantly as pressure increases.

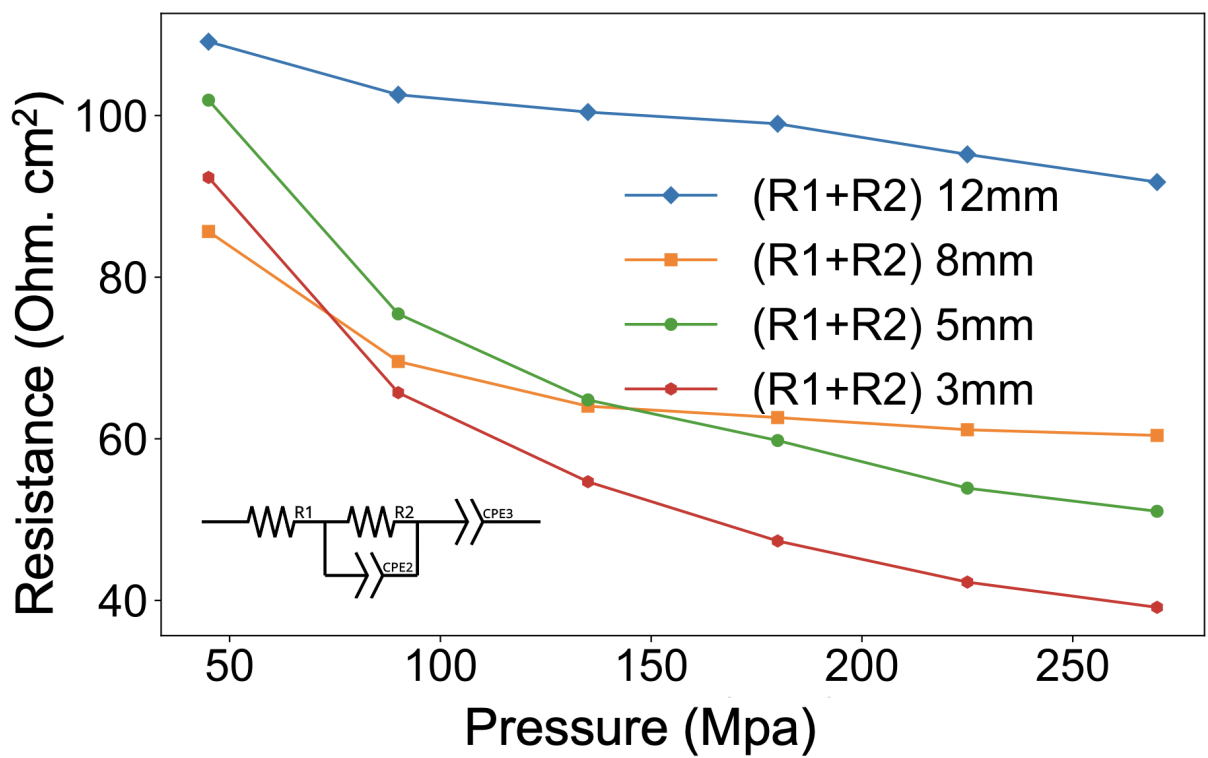

Figure S5: Plot of total resistance (R1 + R2) versus pressure for different electrode contact diameters (12 mm, 8 mm, 5 mm, and 3 mm), showing that resistance decreases with increasing pressure and contact area. The inset shows the equivalent circuit model used for the analysis.

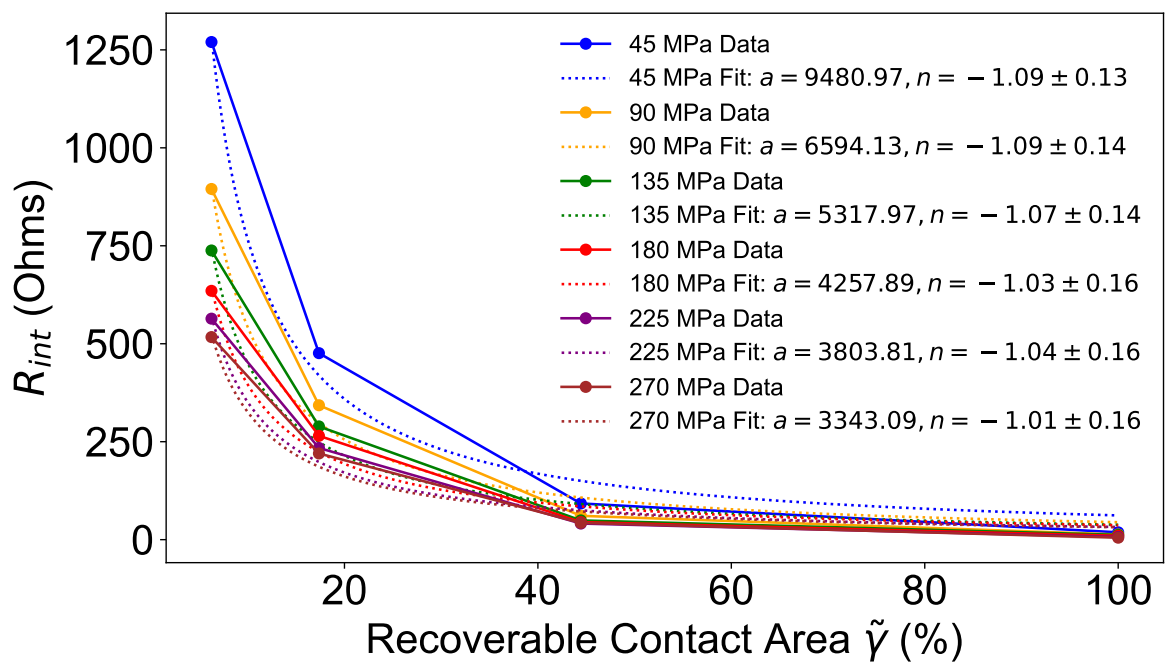

Figure S6: Relationship between interfacial resistance ($R_{\text{int}}$) and recoverable contact area ($\tilde{\gamma}$) for different applied pressures. Experimental data (●) and corresponding power-law fits ($R_{\text{int}} = a\tilde{\gamma}^n$) are shown. The fit parameters ($a$ and $n$) with uncertainties for n are provided for each pressure. The consistent $n$ values across pressures emphasize the robustness of the power-law scaling with recoverable contact area.

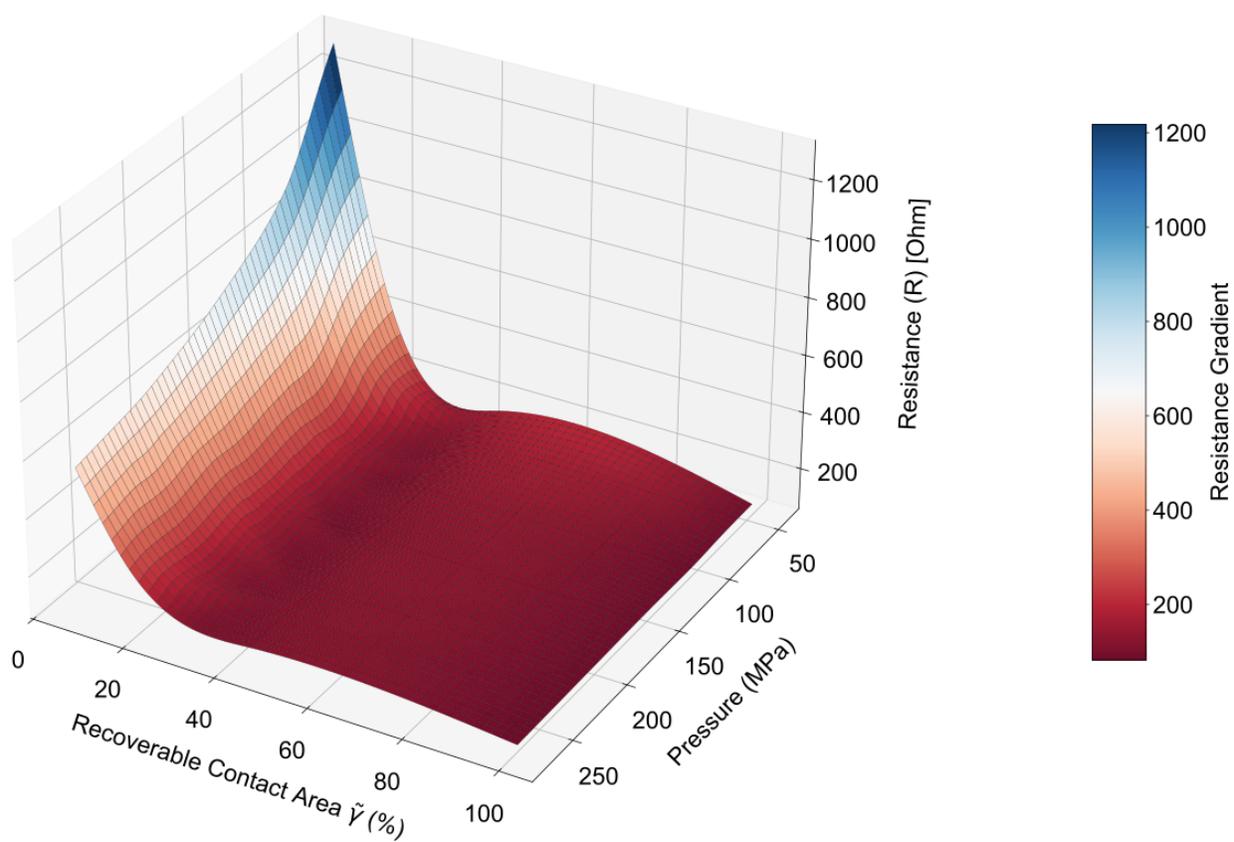

Figure S7: 3D surface map illustrating the variation of resistance ($R$) as a function of pressure ($P$) and recoverable contact area ($\tilde{\gamma}$).

# Finite-element model details

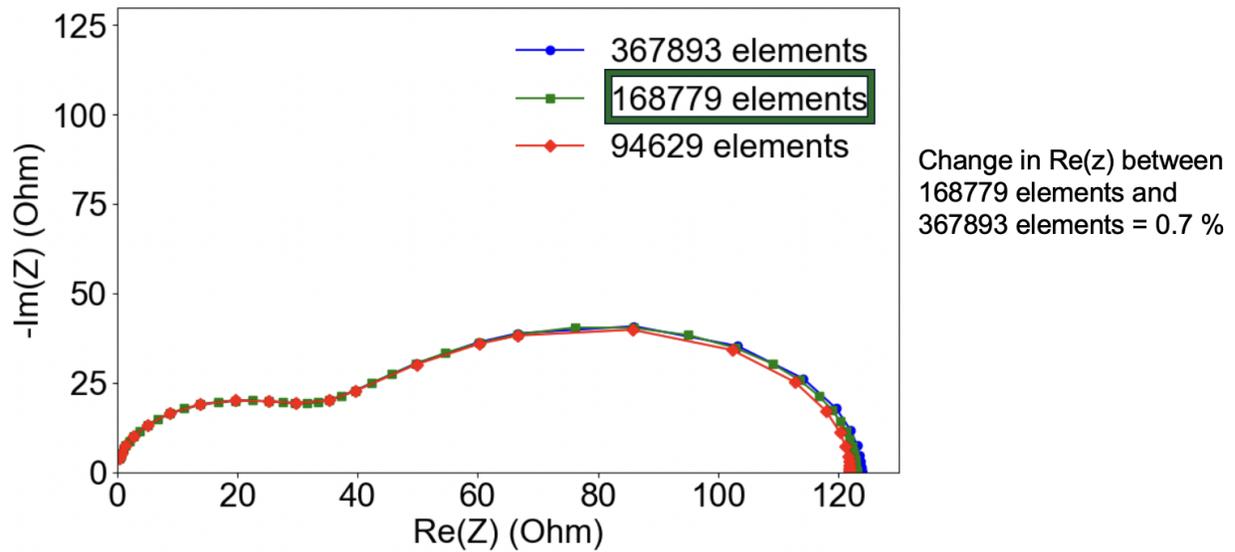

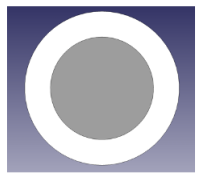

1 x ⌀ 5.9 mm

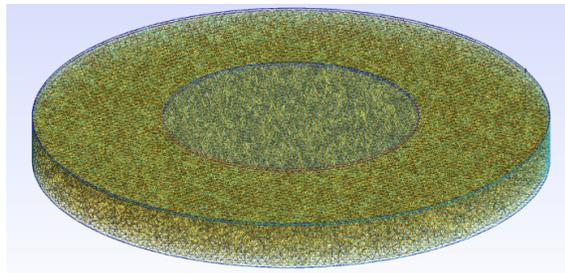

The mesh dependency test demonstrates the convergence of simulation results with respect to mesh size, with only a 0.7% change in total resistance observed between finer meshes. Based on this, the chosen mesh provides a balance between computational efficiency and accuracy.

Figure S8: Mesh dependency test demonstrating simulation convergence, with only a 0.7% change in resistance between finer meshes.

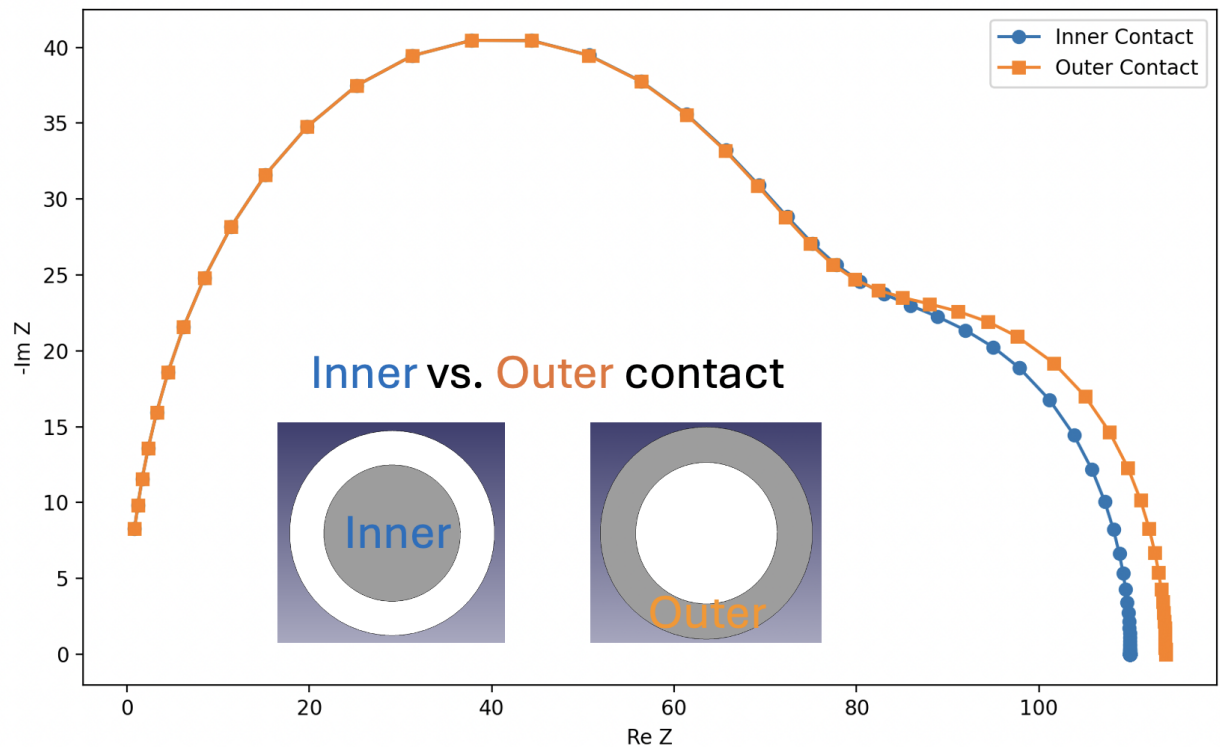

Figure S9: Comparison of simulation results for inner circular and outer annular contacts with equal contact areas, showing that the inner circular contact exhibits lower resistance than the outer annular contact. Hence, our simulations suggest that establishing central contact is more beneficial.

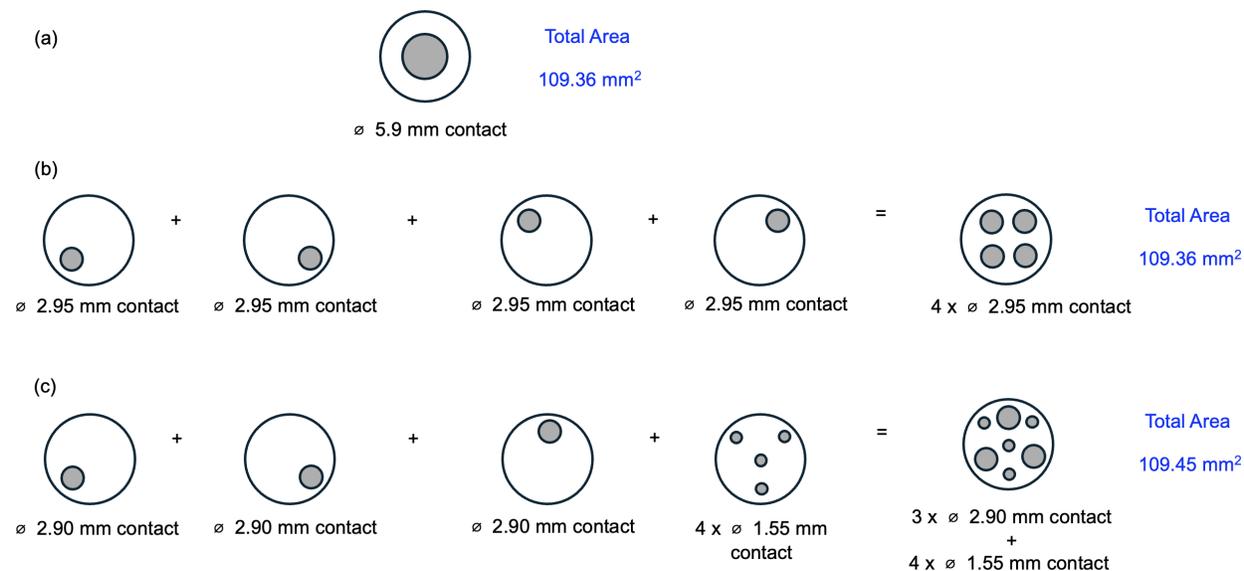

Figure S10: Illustration of the three cases of same contact area analysis with varying geometric distributions.

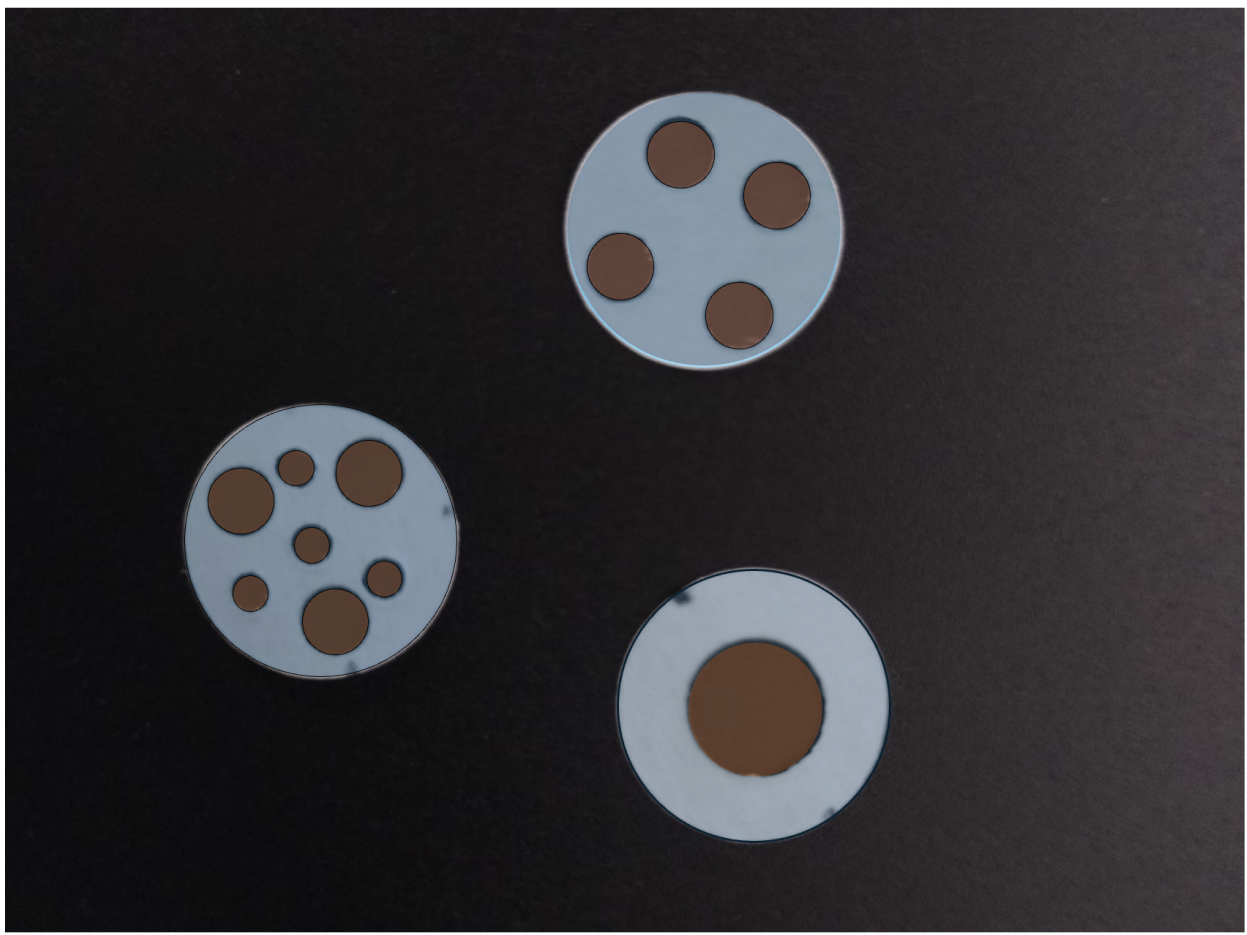

Figure S11: Experimental paper templates for the three same total contact area cases. The white portions represent the paper, and the brown circles indicate the holes allowing electrode-electrolyte contact.

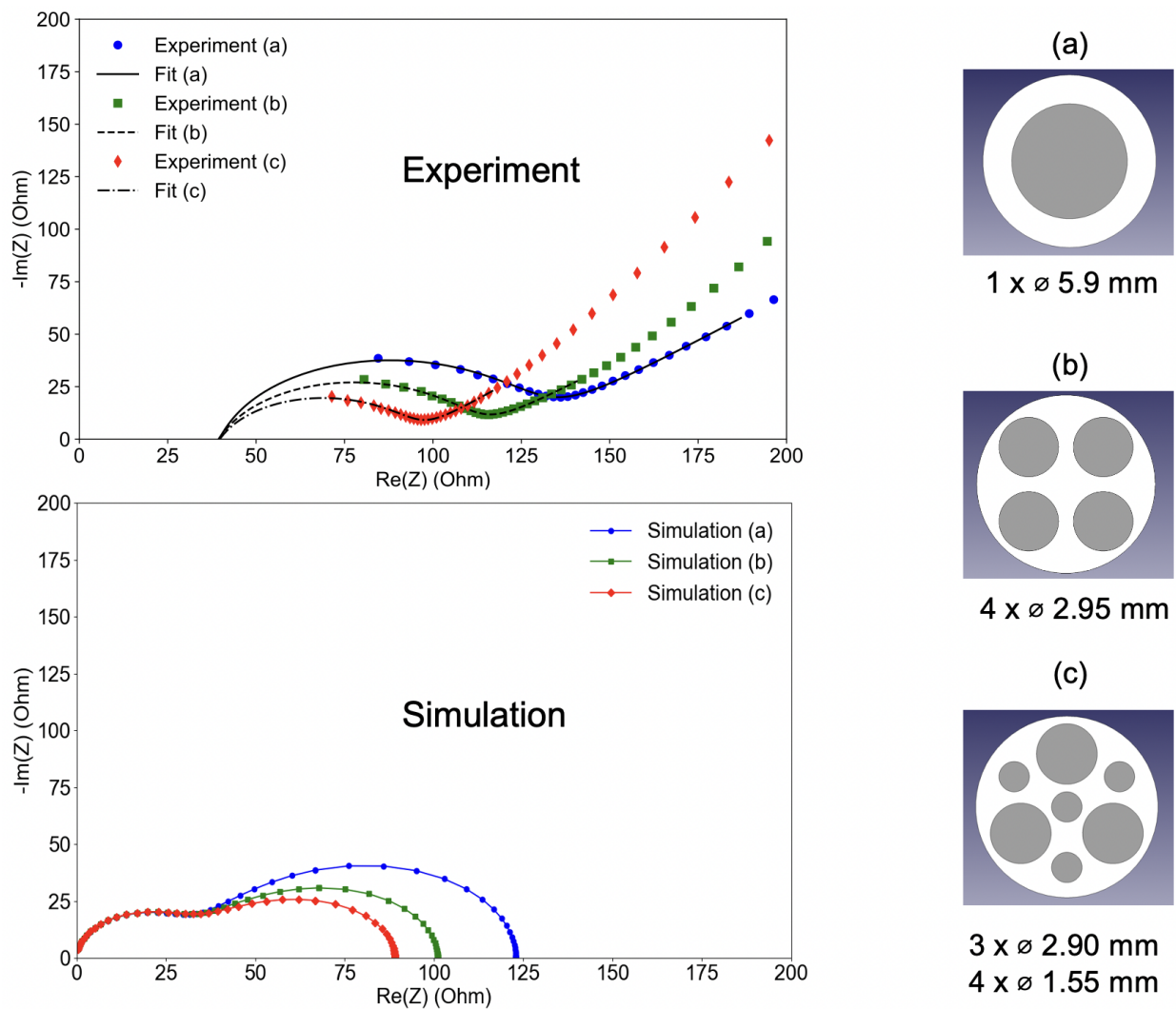

Figure S12: Experimental EIS plots with their fits for the three same total area cases (top) and simulated EIS plots for the corresponding cases (bottom). This figure demonstrates the effect of distributed contacts on reducing interfacial resistance and highlights the strong agreement between experimental results and simulations.

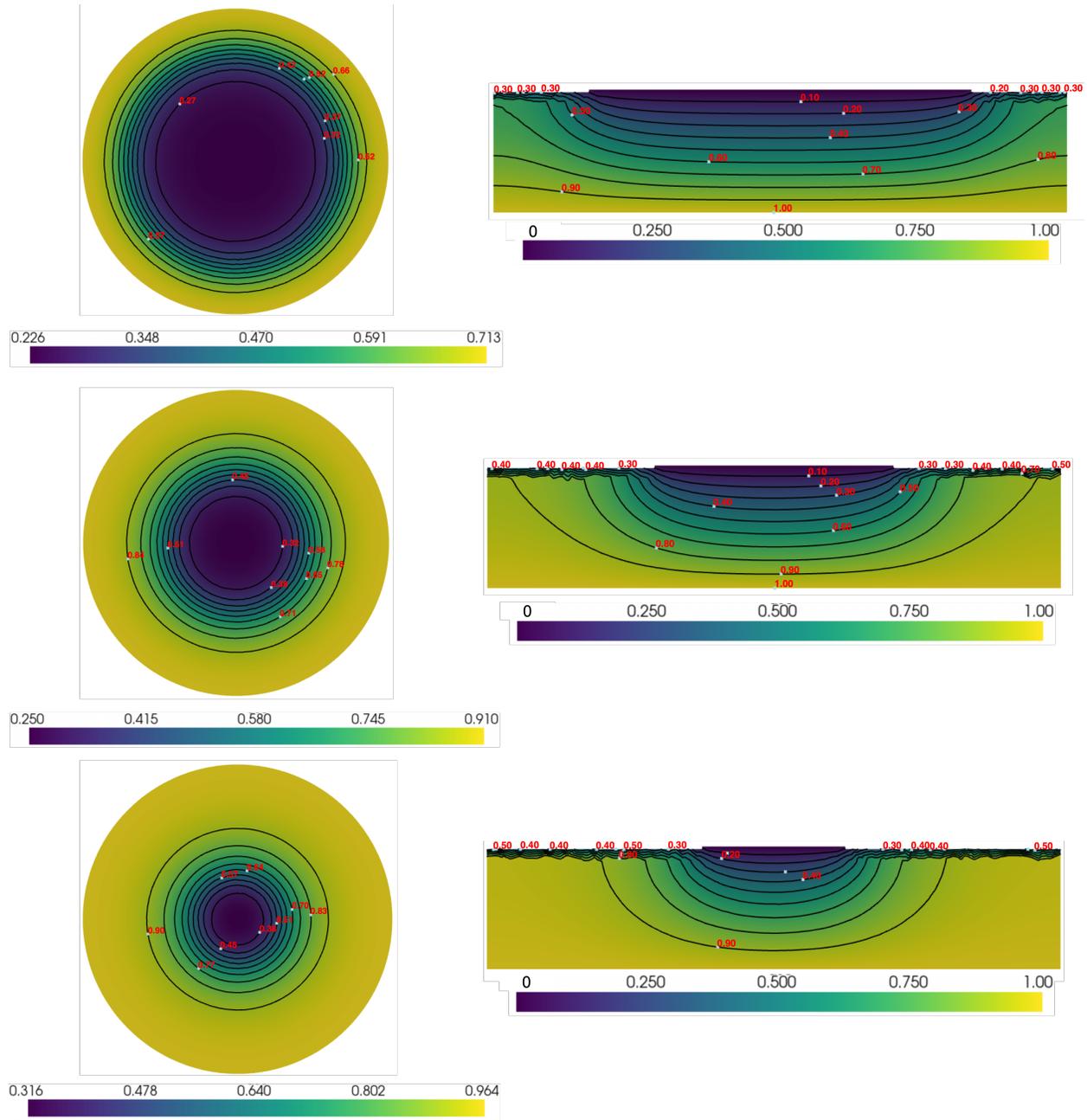

Figure S13: Potential distribution for the three cases of different contact area analysis.

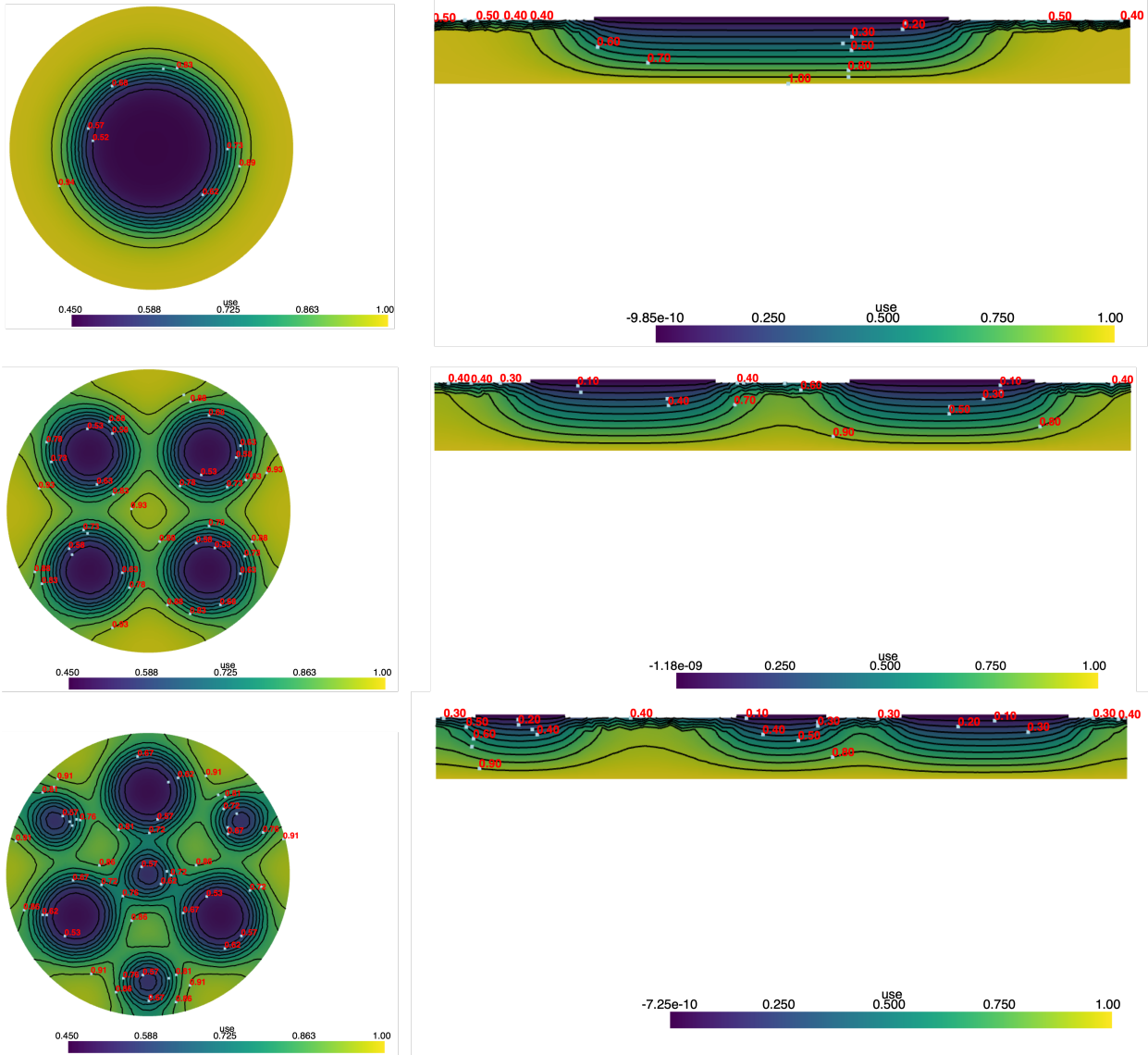

Figure S14: Potential distribution for the three cases of same contact area analysis with varying geometric distributions.

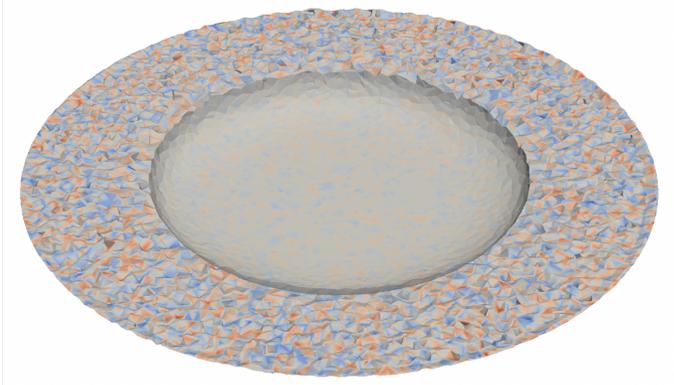

**Isosurfaces illustrating the distribution of the same potential across three different contact geometries with the same total contact area. Despite the identical total contact area, the potential distribution varies significantly, with more distributed geometries showing a more uniform isosurface.**

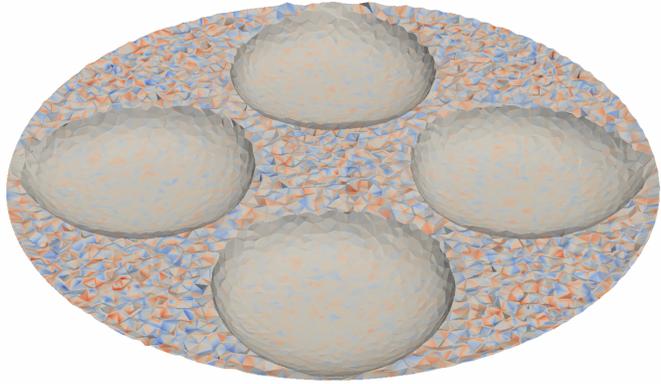

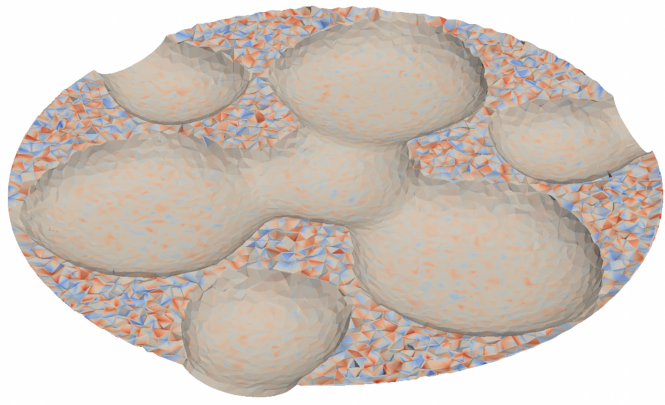

Figure S15: Isosurfaces illustrating potential distribution at the electrode/electrolyte interface. The potential drop is rapid at contact areas, while it remains low in regions of contact loss.

**Different Contact Area Results:**

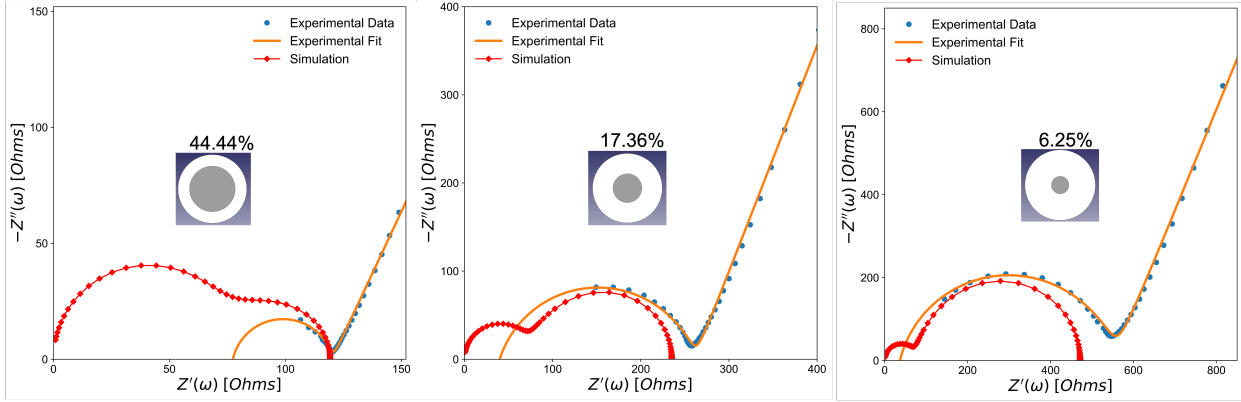

**Same Contact Area Results:**

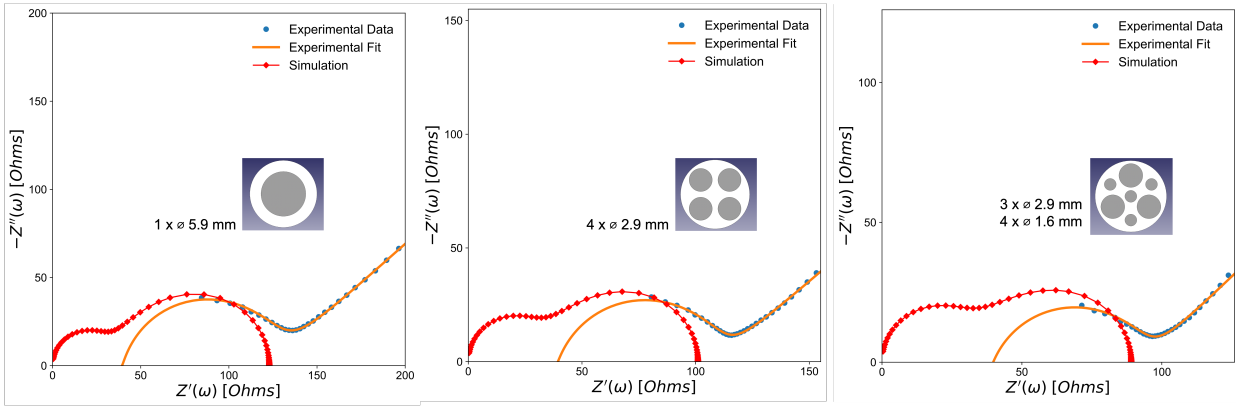

Figure S16: Comparison of experimental and simulation results for different contact area (top row) and same contact area (bottom row) configurations. Top row: Results for varying contact areas with percentage labeled. Bottom row: Results for equal contact areas using different geometrical arrangements (single circular, four smaller circular, and a combination of three and four smaller circular contacts).


\end{document}